\documentclass[12pt]{article}

\usepackage[totalheight = 23cm, totalwidth = 17cm]{geometry}
\usepackage{amssymb,amsmath,amsfonts,amsbsy,epsfig,psfrag}

\begin{document}

\title{Sustainability of multi-field inflation\\and bound on string scale}

\author{
Jinn-Ouk Gong\footnote{jgong@hri.res.in}
\\ \\
\textit{Harish-Chandra Research Institute}
\\
\textit{Chhatnag Road, Jhunsi}
\\
\textit{Allahabad 211 019}
\\
\textit{India} }

\date{\today}

\maketitle

\begin{abstract}

We study the effects of the interaction terms between the inflaton fields on the
inflationary dynamics in multi-field models. With power law type potential and
interactions, the total number of $e$-folds may get considerably reduced and can
lead to unacceptably short period of inflation. Also we point out that this can
place a bound on the characteristic scale of the underlying theory such as string
theory. Using a simple multi-field chaotic inflation model from string theory, the
string scale is constrained to be larger than the scale of grand unified theory.
Allowing post-inflationary generation of perturbation can greatly alleviate this
constrain.

\end{abstract}

\thispagestyle{empty}
\setcounter{page}{0}
\newpage
\setcounter{page}{1}

\section{Introduction}

Although inflation~\cite{inflation1,inflation2,inflation3} is considered to be able
to solve many cosmological problems while providing the desirable initial conditions
for the hot big bang universe, the practical implementation of an inflationary
scenario consistently in the context of high energy theory and phenomenology is
still unclear~\cite{neutrino}. One of the most profound difficulties is the
realisation of the simple chaotic inflation with power law potential~\cite{chaotic},
which fits best with most recent cosmological
observations~\cite{obspowerlawinf1,obspowerlawinf2}. This simplest possibility
typically requires an initial value of the inflaton field far beyond the Planck
scale $m_\mathrm{Pl} \approx 2.4 \times 10^{18}\mathrm{GeV}$. This super-Planckian
initial value exposes any explicit model of chaotic inflation to uncontrollable
radiative corrections and the inflationary predictions become not reliable.

Fortunately, this problem can be evaded in multi-field inflation models, since the
Hubble parameter $H$ receives contributions from all the fields participating in
inflation~\cite{assisted1,assisted2}. Thus even with sub-Planckian field values
which for the single field case will not give rise to enough number of $e$-folds
$\mathcal{N}$, the duration of inflation can be long enough to achieve all the major
successes we expect well under the theoretical control. Moreover, there exist plenty
of scalar fields in theories beyond the standard model of particle physics such as
supersymmetry and supergravity~\cite{susysugra}. Hence any inflationary model based
on such a theory would naturally incorporate multiple number of fields\footnote{The
inflaton candidates are also expected to carry the standard model charges to
populate the observed particle species, especially in the context of minimal
supersymmetric standard model~\cite{mssminf}.}.

At this point, the existence of the coupled terms between the inflaton fields is of
great importance for the inflationary dynamics. In the single field case, the
interactions with most non-inflaton fields are usually required to be very small to
maintain the sufficiently flat inflaton potential. In multi-field models, however,
virtually all the light scalar fields may contribute and there is no reason to
suppress the coupled terms a priori. Also, any phenomenological constraint for the
interaction from such a consideration should place an useful, important bound on the
parameters of the underlying theory. This is what we would like to study in the
present note: we will address this issue using a rudimentary model of multi-field
inflation, introducing a simple cross term which couples the inflaton fields. The
potential may arise in the context of string theory by breaking the shift symmetry,
and the interaction terms appear with their magnitude being suppressed by the ultra
violet cutoff scale, i.e. the string scale $M_s$. A simple analysis on the
inflationary dynamics can place a bound on this scale.

\section{Inflation with coupled terms}

For clarity, we first concentrate on the simplest case of the two-field chaotic
inflation~\cite{doubleinflation1,doubleinflation2} with the potential
\begin{equation}\label{V}
V_\mathrm{leading} = \frac{1}{2}m_\phi^2\phi^2 + \frac{1}{2}m_\chi^2\chi^2 \, ,
\end{equation}
and now we include an interaction term\footnote{Note that a negative sign would
induce a tachyonic instability and may lead to hybrid inflation~\cite{hybrid} and
even dark energy~\cite{hybridde}.}
\begin{equation}\label{interaction}
V_\mathrm{int} = \frac{1}{2}g^2\phi^2\chi^2 \, ,
\end{equation}
where $g$ is the coupling between $\phi$ and $\chi$. In addition, we take $m_\phi$
larger than $m_\chi$ and assume that $m_\phi$ and $m_\chi$ are not too different: if
the difference is too large, the inflationary phase is completely dominated by
$\phi$ and virtually $\chi$ is decoupled from the dynamics, although after inflation
it may serve as a curvaton candidate~\cite{curvaton1,curvaton2}, adding additional
constraints on e.g. the energy scale of
inflation~\cite{curvatoncondition1,curvatoncondition2}.

\begin{figure}[h]
\begin{center}
\epsfig{file=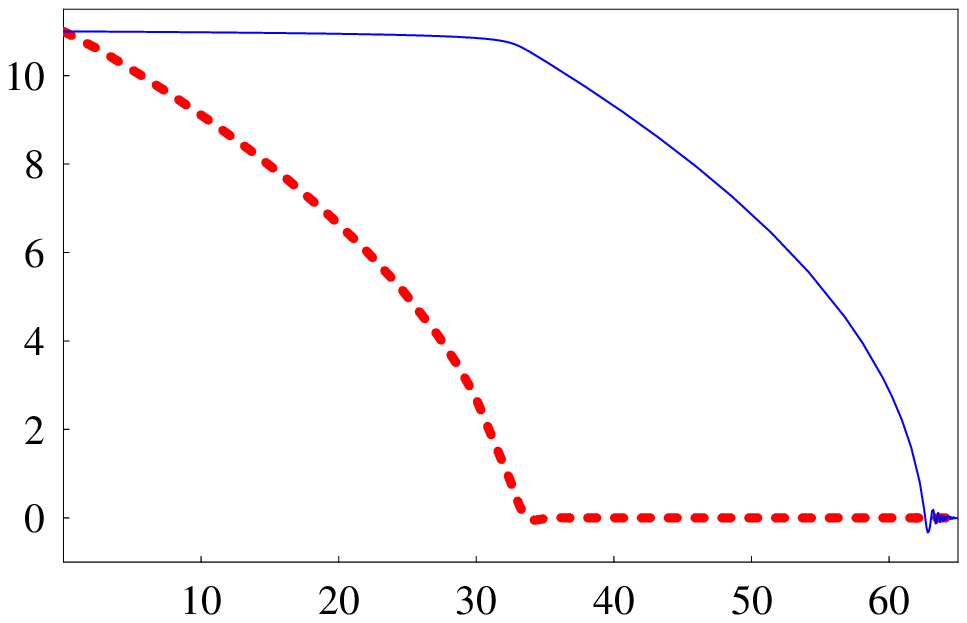, width = 7cm}%
\epsfig{file=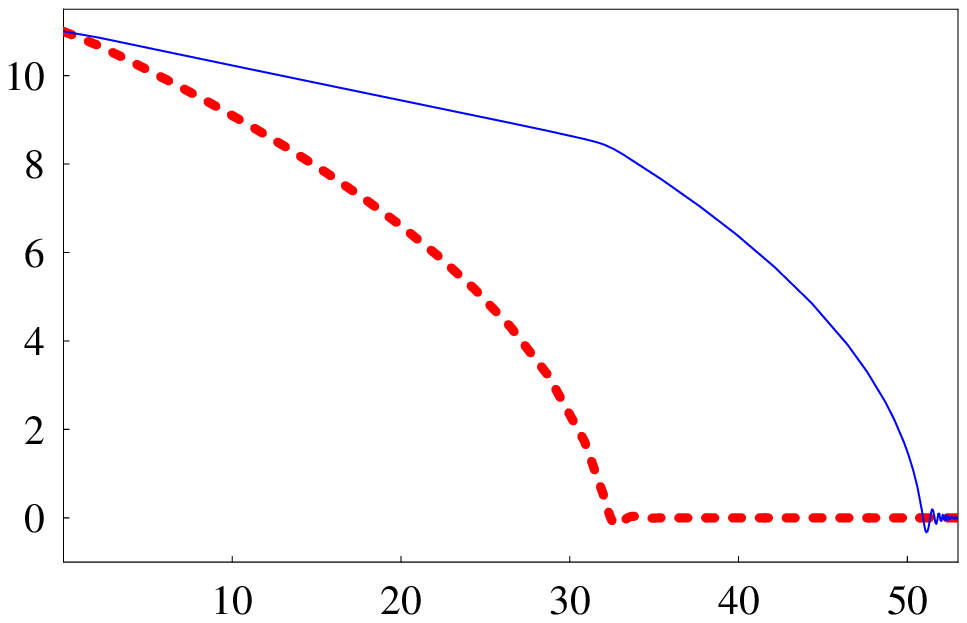, width = 7cm}%
\end{center}
\begin{center}
\epsfig{file=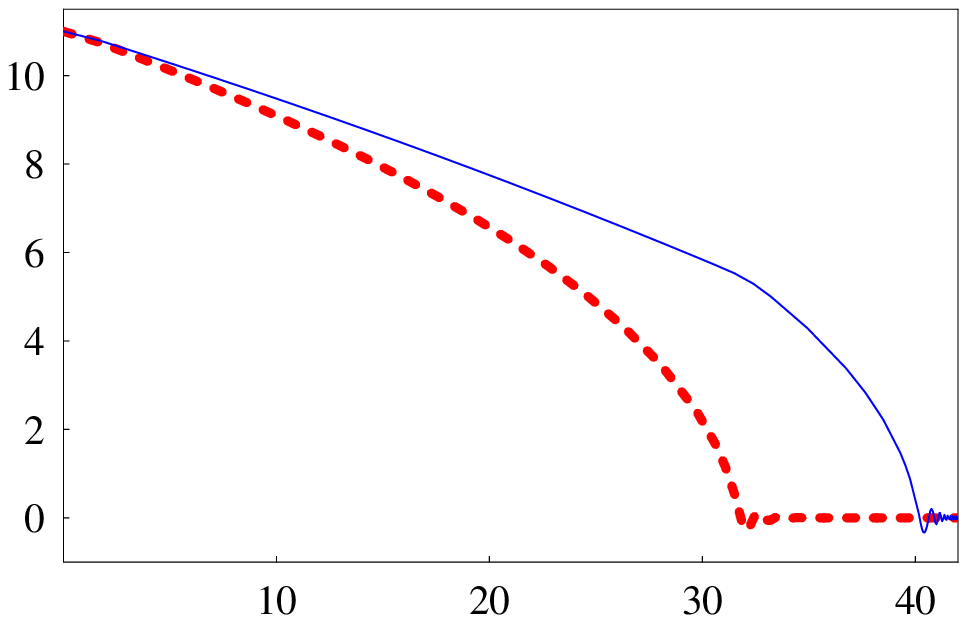, width = 7cm}%
\epsfig{file=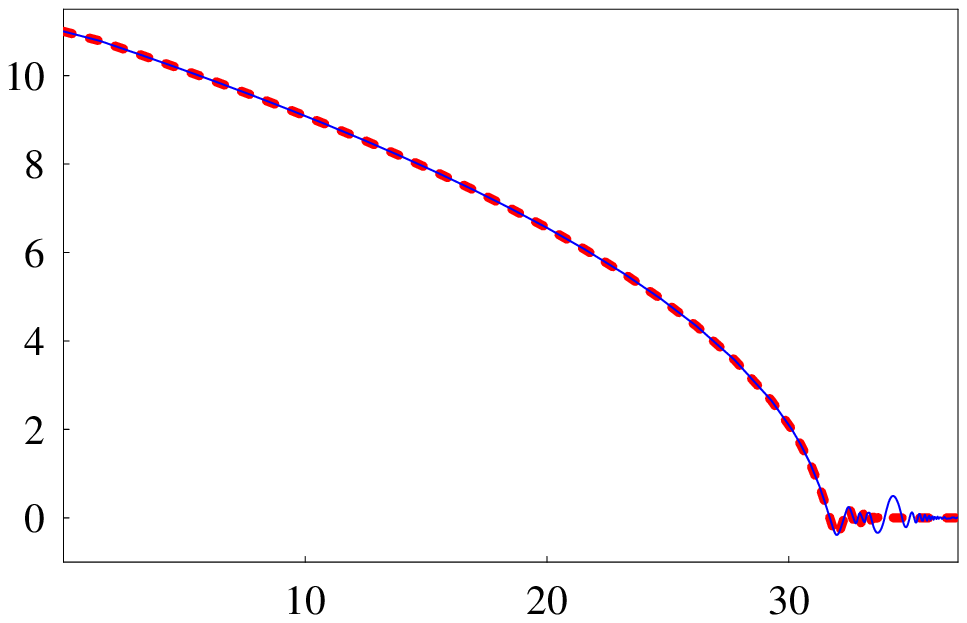, width = 7cm}%
\end{center}
\caption{The evolution of $\phi$ (thick dotted line) and $\chi$ (thin solid line)
versus $\mathcal{N}$. From the upper left plot, we have set $g = 0$, $8 \times
10^{-7}$, $2 \times 10^{-6}$ and $5 \times 10^{-4}$. We set the initial values
$\phi_0 = \chi_0 = 11 m_\mathrm{Pl}$, which without Eq.~(\ref{interaction}) would
give $\mathcal{N} \approx 60$ from Eq.~(\ref{efoldquadratic}). The bare masses are
set to be $m_\phi = 10^{-5}m_\mathrm{Pl}$ and $m_\chi = m_\phi/10$, which gives
$\mathcal{P}^{1/2} \approx 5.58696 \times 10^{-5}$ at 60 $e$-folds before the end of
inflation. Note that with large number of the fields participating in the
inflationary dynamics, significant curvature perturbation may be generated after
inflation with the observed spectral index~\cite{afterinf} as well as non-negligible
isocurvature one~\cite{choi}. The masses then can be quite light, but $\mathcal{N}$
still remains more or less the same. The total number of $e$-folds until the end of
inflation is, again from the upper left plot, 62.0865, 50.3970, 39.6637 and 30.5004,
respectively.}
\label{num}
\end{figure}

In Fig.~\ref{num}, we present several numerical results. As can be seen, when the
interaction term Eq.~(\ref{interaction}) dominates, the total number of $e$-folds is
reduced almost by half. More importantly, when the coupling is strong $\phi$ and
$\chi$ tend to follow the same evolution, i.e. in the field space the trajectory is
directly towards the minimum at the origin. Also note that for $g \gtrsim 5 \times
10^{-6}$, $\mathcal{N}$ hardly changes.

We can understand the evolution of $\phi$ and $\chi$ as follows. Given the potential
Eq.~(\ref{V}) with the interaction term Eq.~(\ref{interaction}), the equations of
motion of $\phi$ and $\chi$ are written as
\begin{align}\label{eomphi}
\ddot\phi + 3H\dot\phi + \left( m_\phi^2 + g^2\chi^2 \right)\phi = & 0 \, ,
\\
\ddot\chi + 3H\dot\chi + \left( m_\chi^2 + g^2\phi^2 \right)\chi = & 0 \, ,
\label{eomchi}
\end{align}
respectively, where the Hubble parameter $H$ during inflation is given by
\begin{equation}
H^2 = \frac{1}{3m_\mathrm{Pl}^2} \left( \frac{1}{2}\dot\phi^2 +
\frac{1}{2}\dot\chi^2 + V \right) \, ,
\end{equation}
with $V$ being the sum of Eqs.~(\ref{V}) and (\ref{interaction}). From
Eqs.~(\ref{eomphi}) and (\ref{eomchi}), we can see that the ``effective'' masses are
composed of the bare masses and the interaction, i.e.
\begin{align}\label{mphieff}
m_{\phi(\mathrm{eff})}^2 = & m_\phi^2 + g^2\chi^2 \, ,
\\
m_{\chi(\mathrm{eff})}^2 = & m_\chi^2 + g^2\phi^2 \, ,
\end{align}
for $\phi$ and $\chi$, respectively. Thus two different effects are competing, and
the dynamics is determined by the term which is important: if the bare masses
dominate, $\phi$ and $\chi$ are practically decoupled and it is well known that the
total number of $e$-folds $\mathcal{N}$, neglecting $\mathcal{O}(1)$ corrections, is
given by~\cite{doubleinflation1,doubleinflation2}
\begin{equation}\label{efoldquadratic}
\mathcal{N} \equiv \int H dt = \frac{\phi_0^2 + \chi_0^2}{4m_\mathrm{Pl}^2} \, .
\end{equation}
Conversely, when the interaction is strong enough, this prediction becomes
different: the equations of motion, especially at the early stage of inflation, are
simplified to, with the replacement $\rho^2 = \phi^2 + \chi^2$,
\begin{equation}\label{2fieldeomsimple}
\ddot\rho + 3H\dot\rho + g^2\rho^3 = 0 \, ,
\end{equation}
i.e. the equation becomes that of the quartic potential and both $\phi$ and $\chi$
essentially follow the same evolution. In this case, $\mathcal{N}$ is given by
\begin{equation}\label{quarticN}
\mathcal{N} = \frac{\phi_0^2 + \chi_0^2}{8m_\mathrm{Pl}^2} \, ,
\end{equation}
and is reduced by half compared with the case where the bare masses are dominating,
which is clear from Fig.~\ref{num}. This is also why the coupling larger than a
specific value does not alter $\mathcal{N}$ appreciably: this value corresponds to
what makes the interaction term larger than the bare mass contribution.

\begin{figure}[h]
\psfrag{N}{$\mathcal{N}$}%
\psfrag{logx}{$g$}%
\psfrag{logP}{$\mathcal{P}^{1/2}$}%
\begin{center}
\epsfig{file = 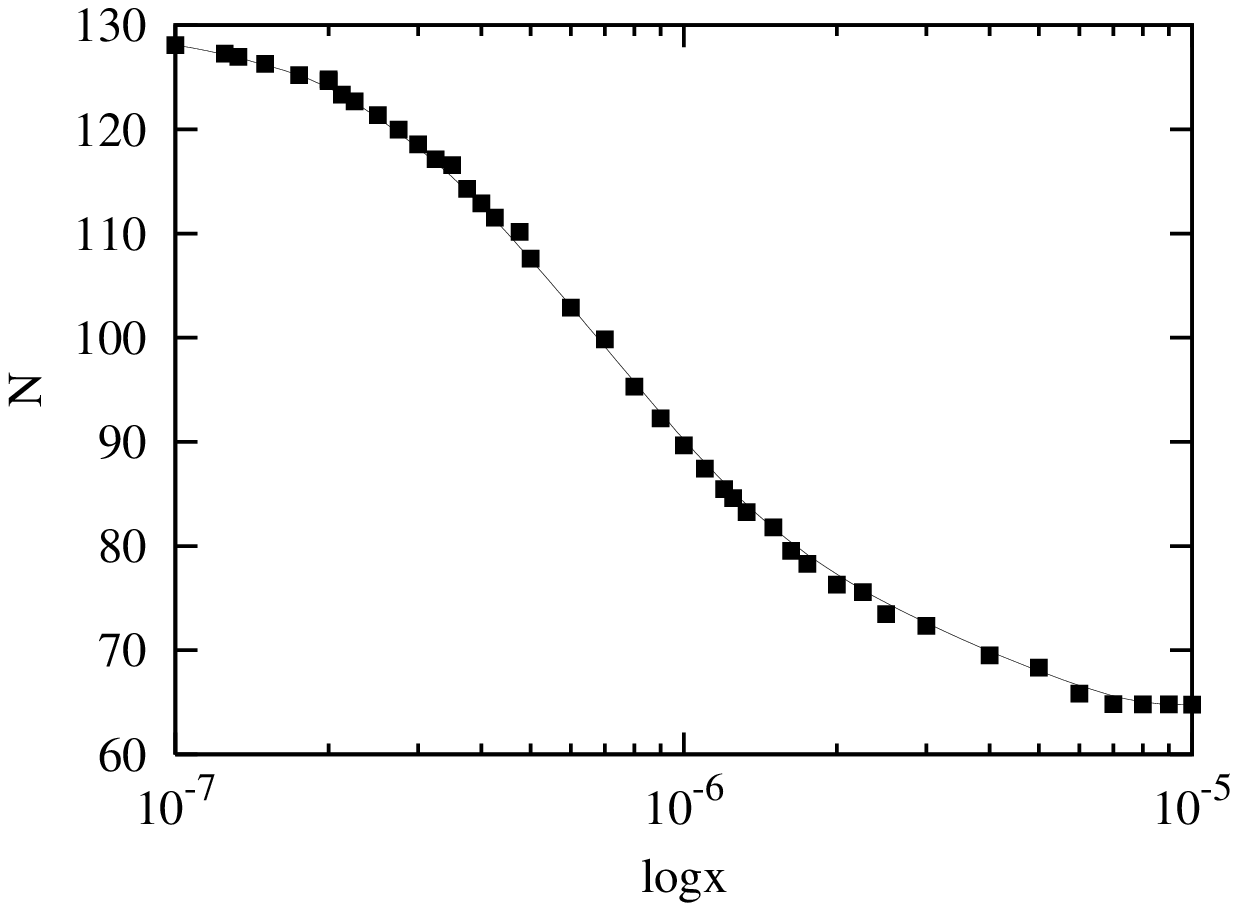, width = 6cm, angle = -90}%
\epsfig{file = 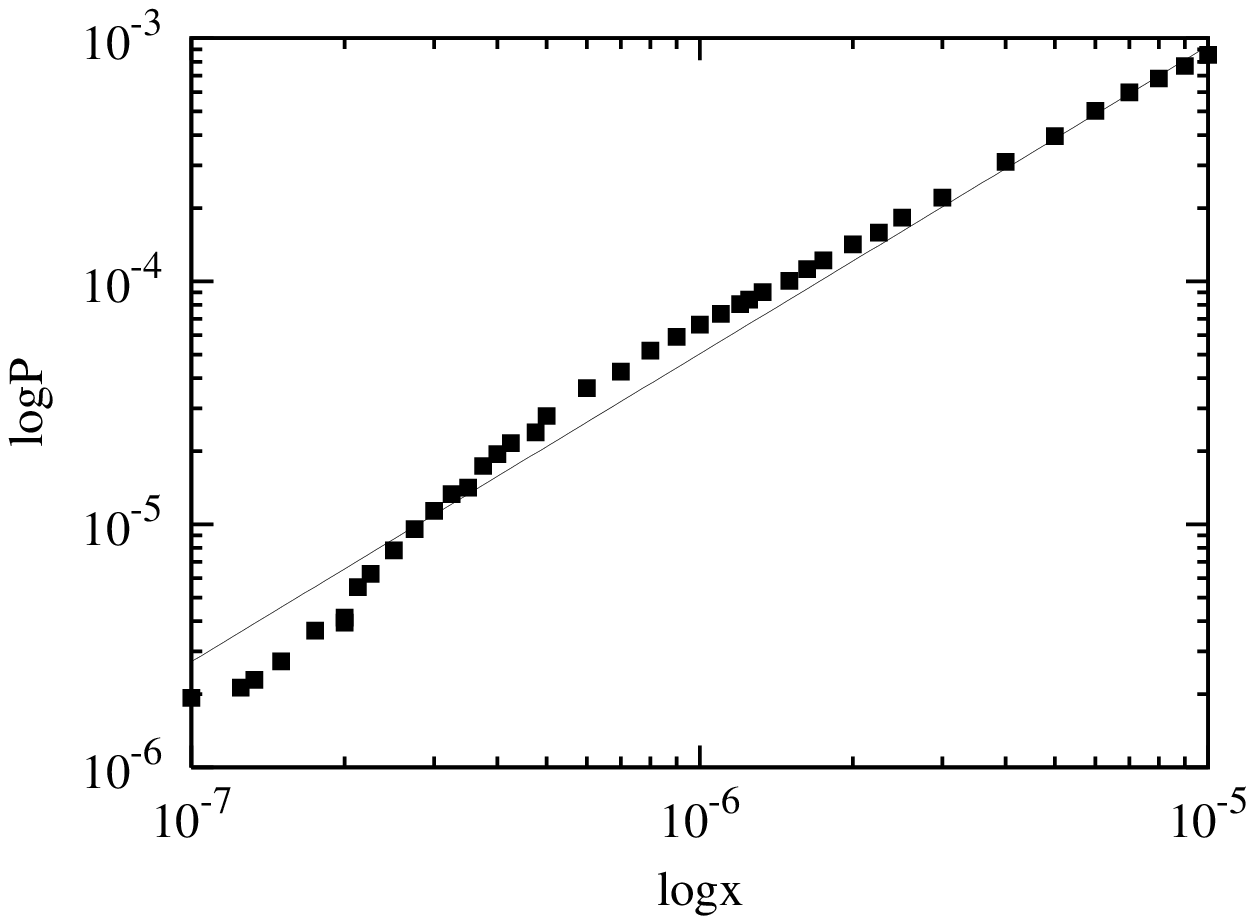, width = 6cm, angle = -90}
\end{center}
\caption{The plots of (left) $\mathcal{N}$ and (right) $\mathcal{P}^{1/2}$ versus
$g$. Square dots denote numerical results, and solid lines are interpolated curves.
We have used the same masses given in Fig.~\ref{num}, but this time the initial
values are set to be $\phi_0 = \chi_0 = 16m_\mathrm{Pl}$ so that even when the
interaction completely dominates the potential, with $\mathcal{N}$ given by
Eq.~(\ref{quarticN}), we still have more than 60 $e$-folds. In the right panel, we
have evaluated $\mathcal{P}^{1/2}$ at 60 $e$-folds before the end of inflation. As
can be seen in the left panel, $\mathcal{N}$ is reduced almost by half when the
interaction becomes strong. Note that for a wide range of $g$, $\mathcal{N}$ remains
more or less the same and only the value of $g$ taken from a limited range can
significantly modify $\mathcal{N}$. The constraint on $g$ beyond this range can be
obtained from the magnitude of $\mathcal{P}^{1/2}$, which is approximately $5 \times
10^{-5}$ by recent observations~\cite{obspowerlawinf1,obspowerlawinf2}.
$\mathcal{P}^{1/2}$ is linearly dependent on $g$ in logarithmic scale, which
suggests that Eq.~(\ref{Pquartic}) is indeed the case. According to the above
numerical results, $g \sim 10^{-6}$ gives the observed value of $\mathcal{P}^{1/2}$
with the effect of the interaction being prominent but not completely dominating:
see the lower left plot of Fig.~\ref{num}. We have also calculated the spectral
index $n_\mathrm{s}$, but irrespective of the value of $g$ it remains between 0.95
and 0.97. From Eqs.~(\ref{indexquardratic}) and (\ref{indexquartic}), it is expected
that $n_\mathrm{s}$ is almost independent of $g$.} \label{gdependence}
\end{figure}

The case of intermediate interacting energy should be of most interest. Since the
slow roll approximation is valid for most duration, we can write the equations of
motion for $\phi$ and $\chi$, in terms of $\mathcal{N}$, as
\begin{equation}\label{sreom}
\phi_i' + m_\mathrm{Pl}^2 \frac{V_{,i}}{V} = 0 \, ,
\end{equation}
where we have used $H^2 \approx V/(3m_\mathrm{Pl}^2)$, a prime denotes a derivative
with respect to $\mathcal{N}$ and the subscript $i$ stands for $\phi$ and $\chi$. We
can thus write the solution of Eq.~(\ref{sreom}) as
\begin{equation}\label{srsol}
\phi_i \approx \phi_{i(0)} - \sqrt{2\epsilon_i}\mathcal{N}m_\mathrm{Pl} \, ,
\end{equation}
where the slow-roll parameter
\begin{equation}\label{srepsilon}
\epsilon_i \equiv \frac{m_\mathrm{Pl}^2}{2} \left( \frac{V_{,i}}{V} \right)^2
\end{equation}
is assumed to be nearly constant, which is a good enough approximation in the slow
roll regime of inflation. Since we are assuming that the energy associated with the
heavier field, $\phi$ here, and the interaction energy are relatively large, the
number of $e$-folds until $\phi$ contributes follows that of a quadratic potential
of $\phi$ which is independent of the effective mass of $\phi$ given by
Eq.~(\ref{mphieff}), and is estimated simply as
\begin{equation}\label{Nphi}
\mathcal{N}_\phi \approx \frac{\phi_0^2}{4m_\mathrm{Pl}^2} \, .
\end{equation}
Until then the lighter field, $\chi$, evolves according to Eq.~(\ref{srsol}). The
amplitude of $\chi$ at the moment when $\phi$ exits inflationary regime,
$\chi_\phi$, is thus given by
\begin{equation}
\chi_\phi \approx \chi_0 - \frac{\sqrt{2\epsilon_\chi}}{4}
\frac{\phi_0^2}{m_\mathrm{Pl}} \, .
\end{equation}
Afterwards only $\chi$ drives inflation. Hence the number of $e$-folds
$\mathcal{N}_\chi$, achieved during this phase, is simply given by the standard
result
\begin{equation}\label{Nchi}
\mathcal{N}_\chi \approx \frac{\chi_\phi^2}{4m_\mathrm{Pl}^2} \, ,
\end{equation}
where we have assumed for simplicity that inflation ends when $\chi = 0$. The total
number of $e$-folds is then, from Eqs.~(\ref{Nphi}) and (\ref{Nchi}),
\begin{align}\label{totalN}
\mathcal{N} & \approx \mathcal{N}_\phi + \mathcal{N}_\chi
\nonumber\\
& \approx \frac{1}{4m_\mathrm{Pl}^2} \left[ \phi_0^2 + \left( \chi_0 -
\frac{\sqrt{2\epsilon_\chi}}{4} \frac{\phi_0^2}{m_\mathrm{Pl}} \right)^2 \right] \,
.
\end{align}
For example, let us take the numbers from the upper right plot of Fig.~\ref{num}.
With the given numbers, roughly
\begin{align}\label{Eorder}
\frac{1}{2}m_\phi^2\phi_0^2 \sim 10^{-9}m_\mathrm{Pl}^4 \, ,
\nonumber\\
\frac{1}{2}m_\chi^2\chi_0^2 \sim 10^{-11}m_\mathrm{Pl}^4 \, ,
\nonumber\\
\frac{1}{2}g^2\phi_0^2\chi_0^2 \sim 10^{-9}m_\mathrm{Pl}^4 \, .
\end{align}
Thus, from Eq.~(\ref{Nphi}), $\mathcal{N}_\phi \approx 30$. Meanwhile, using
Eq.~(\ref{srepsilon}), we can estimate
\begin{align}\label{epsilonchi}
\sqrt{2\epsilon_\chi} & = 2m_\mathrm{Pl} \frac{(m_\chi^2 +
g^2\phi_0^2)\chi_0}{m_\phi^2\phi_0^2 + m_\chi^2\chi_0^2 + g^2\phi_0^2\chi_0^2}
\nonumber\\
& \approx g^2 \frac{\chi_0 m_\mathrm{Pl}}{m_\phi^2}
\nonumber\\
& \approx 0.07 \, ,
\end{align}
where we have used our assumption that the potential energy of $\phi$ and the
interaction energy are of the same order of magnitude and are very large compared
with the potential energy of $\chi$, as can be read from Eq.~(\ref{Eorder}).
Substituting Eq.~(\ref{epsilonchi}) into Eq.~(\ref{totalN}), then we have
\begin{equation}
\mathcal{N} \approx 50 \, ,
\end{equation}
which is in good agreement with the numerical estimate $\mathcal{N} \approx
50.3970$. Also we note that both the bare masses and the coupling do not affect the
duration of the inflationary epoch but they determine the amplitude of the power
spectrum $\mathcal{P}$, see Eqs.~(\ref{Pquadratic}) and (\ref{Pquartic}). In
Fig.~\ref{gdependence}, we show the dependence of $\mathcal{N}$ and $\mathcal{P}$ on
the coupling $g$.

We can easily expand the argument to include arbitrary number of fields: there are
total $N$ fields $\phi_i$, each with the mass $m_i$ so that the leading piece is
\begin{equation}\label{generalV}
V_\mathrm{leading} = \sum_{i=1}^N \frac{1}{2}m_i^2\phi_i^2 \, ,
\end{equation}
with the interaction between $\phi_i$ and $\phi_j$ $(i \neq j)$ being given by
\begin{equation}\label{generalinteraction}
V_{ij} = \frac{1}{2}g_{ij}^2\phi_i^2\phi_j^2 \, .
\end{equation}
The potential is then written as
\begin{equation}\label{totalV}
V = \sum_{i = 1}^N \frac{1}{2}m_i^2\phi_i^2 + \sum_{i \neq j}
\frac{1}{2}g_{ij}^2\phi_i^2\phi_j^2 \, .
\end{equation}
Then the equation of motion of $\phi_i$ is given by
\begin{equation}\label{multieom}
\ddot\phi_i + 3H\dot\phi_i + \left( m_i^2 + \sum_{j \neq i} g_{ij}^2\phi_j^2
\right)\phi_i = 0 \, .
\end{equation}
Thus while the bare mass remains fixed, the number of interactions can become very
large according to the total number of fields, making them completely dominating
unless the couplings are all exceptionally small. For simplicity, let us assume that
all the masses and the coupling constants are the same, and write them as $m$ and
$\widetilde{g}$, respectively. Then with $N \gg 1$ so that $N$ and $N-1$ are more or
less the same, Eq.~(\ref{multieom}) is reduced to
\begin{equation}
\ddot\rho + 3H\dot\rho + \left( m^2 + N\widetilde{g}^2\rho^2 \right) \rho = 0 \, .
\end{equation}
Compared with Eq.~(\ref{2fieldeomsimple}), when the interactions are strong,
$\widetilde{g}$ should be suppressed by a factor of $\sqrt{N}$ to maintain the same
predictions.

\begin{figure}[h]
\begin{center}
\epsfig{file=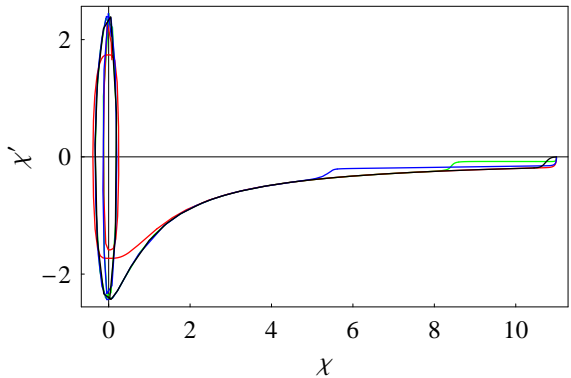, width = 8cm}
\end{center}
\caption{The evolution of $\chi$ in the phase space. The values of the parameters
are taken from Fig.~\ref{num} so that the cases with $g = 0$ (black), $8 \times
10^{-7}$ (green), $2 \times 10^{-6}$ (blue) and $5 \times 10^{-4}$ (red) are shown.
It can be seen that $\chi$ tends to follow an attracting trajectory and that thus
the slow roll approximation is good enough during most duration of evolution. The
`kinks' of the green and blue lines denote the moment of drop out of $\phi$, so that
from then $\chi$ quickly follows the uncoupled evolution.}
\label{phaseplot}
\end{figure}

Now we briefly mention perturbations. We may follow the $\delta\mathcal{N}$
formalism~\cite{deltaN1,deltaN2,deltaN3,deltaN4}, or the standard procedure for the
single field case~\cite{single1,single2}, to calculate the observable quantities
such as the power spectrum $\mathcal{P}$ and the spectral index $n_\mathrm{s}$: if
the masses are dominating, we have
\begin{align}\label{Pquadratic}
\mathcal{P} = & \frac{m^2\rho^4}{96\pi^2m_\mathrm{Pl}^6} \, ,
\\
n_\mathrm{s} - 1 = & -\frac{8m_\mathrm{Pl}^2}{\rho^2} \, ,
\label{indexquardratic}
\end{align}
or when the interactions are important,
\begin{align}\label{Pquartic}
\mathcal{P} = & \frac{g^2\rho^6}{768\pi^2m_\mathrm{Pl}^6} \, ,
\\
n_\mathrm{s} - 1 = & -\frac{24m_\mathrm{Pl}^2}{\rho^2} \, .
\label{indexquartic}
\end{align}
When the masses or the couplings are different, typically we have a redder
spectrum~\cite{redderP}. At this point, it is interesting to note that the results
are confined between those of the $\phi^2$ and $\phi^4$ theories. This can be seen
from the geometry of the field space: the interaction $g_{ij}^2\phi_i^2\phi_j^2$
lifts the whole potential, with the steepest rise along the line $\phi_i = \phi_j$
which corresponds to the quartic potential, while leaves $\phi_i = 0$ and $\phi_j =
0$, i.e. the mass eigenstates, intact. Thus in the field space the descent is
mildest along each axis with quadratic dependence, and is steepest along $\phi_1 =
\phi_2 = \cdots = \phi_N$ with quartic dependence. Thus at the early times the
trajectory tends to be towards the origin along the steepest descent. This geometric
consideration should be applicable to generic power law potential and interactions.
Also we should note that the shape of $\mathcal{P}$ is saw-edged. With not equal
masses (or couplings in general) heavier fields will drop out of the inflationary
regime first. Near the moment of a single drop out, we can expand the potential then
a change of slope is experienced: with $A$ denoting the slope, when $N$-th field is
relaxed to minimum and quits the inflationary regime, we
have~\cite{slopechange1,slopechange2}
\begin{equation}
\left| \frac{\Delta A}{A} \right| \approx \frac{m_N^2\phi_N}{\sum_i m_i^2\phi_i} \,
,
\end{equation}
where the right hand side is evaluated at the moment of the drop out. This will lead
to an enhancement of $\mathcal{P}$ across the corresponding scale, which becomes
mild when the masses are densely spaced. In Fig.~\ref{phaseplot} we plot the phase
portrait of $\chi$ of Fig.~\ref{num}, from which this momentary feature can be read.
It is also noticeable that with strong enough interactions, the level of
non-Gaussianity in general will not be significant. $f_\mathrm{NL}$ during inflation
is known to become large when the trajectory in the field space is highly
non-trivial~\cite{multiNG}. For the case where the interaction terms are dominating,
however, all the fields evolve essentially in the same, or at least similar way,
making the trajectory in the field space varying very smoothly. This will lead to
mild, even negligible enhancement of $f_\mathrm{NL}$. Although it should be possible
to obtain considerable non-Gaussianity with a very special choice of the initial
conditions, but this looks not quite likely.

Finally, we mention the possibility of parametric resonance due to the oscillation
of a field near its minimum. In the theory of
preheating~\cite{preheating1,preheating2}, the strength of parametric resonance is
determined by the parameter
\begin{equation}
q = \frac{g^2\Phi^2}{4m_\phi^2} \, ,
\end{equation}
where $\Phi$ is the amplitude of oscillation of the inflaton $\phi$ when the
inflationary phase ends. Only when $q \gg 1$, the field coupled to $\phi$, say
$\chi$, goes through many instability bands of the Mathieu equation and the
resonance becomes prominent. In the present case, however, because of the relatively
large Hubble parameter, the energy of the massive fields which would oscillate
earlier in the absence of other light fields is dissipated and those fields are
overdamped, showing negligible oscillation. Even for the light fields which
oscillate later, it is known that the amplitude near the end of inflation is
suppressed by a factor proportional to $\sqrt{N}$~\cite{2007feb}, making the
amplitudes of oscillation far smaller than that of the single field case. Hence we
have small $q$ and the resonance is suppressed. If we increase $g$ to make $q$
large, as discussed before all the fields follow the same evolution and oscillate in
phase, heavily suppressing  the possibility of parametric resonance\footnote{There
is, however, some possibility in the context of $\lambda\phi^4$ theory, which is
equivalent to the present situation~\cite{phi4preheating}. However pursuing this
idea is beyond the scope of the present paper, and we do not discuss it here.}. We
have tried different parameter ranges but have not found any instability of $\chi$.
This seems to suggest that we need fine tunings to observe any parametric resonance.

\section{Bound on string scale}

Now we turn to our next point how the discussion in the previous section may put a
bound on the string scale $M_s$. It was argued~\cite{Nflation} that in string theory
the ubiquitous string axion fields can give rise to the uncoupled quadratic
potential, Eq.~(\ref{generalV}), as the leading approximation, provided that the
fields are displaced not too far from their minima. The potential is a sum of the
axion contributions\footnote{The next higher order self-interaction near the minimum
will then be negative, $-(2\pi\Lambda_i/f_i)^4 \phi_i^4/24$. Thus we do not consider
these contributions. But even if we take them into account, still the argument here
is applicable.},
\begin{equation}\label{axionV}
V_\mathrm{leading} = \sum_{i=1}^N \Lambda_i^4 \left[ 1 - \cos \left(
\frac{2\pi\phi_i}{f_i} \right) \right] \, ,
\end{equation}
where $\Lambda_i$ is the axion potential scale and $f_i$ is the axion decay
constant. Multi-instanton corrections generate the coupling terms between the
axions,
\begin{equation}\label{axionVcorrection}
V_{ij} = \frac{\Lambda_i^4\Lambda_j^4}{M^4} \cos \left( \frac{2\pi\phi_i}{f_i}
\right) \cos \left( \frac{2\pi\phi_j}{f_j} \right) \, .
\end{equation}
where $M$ is the cutoff scale of the theory, which in string theory should
correspond to the string scale $M_s$. Typically $\Lambda_i$ is smaller than
$m_\mathrm{Pl}$ by many orders of magnitude thus provided that $\Lambda_i \ll M_s$
Eq.~(\ref{axionVcorrection}) is highly suppressed and can be ignored so that
Eq.~(\ref{generalV}) is a good enough approximation of Eq.~(\ref{axionV}).

For small field values, we can expand Eq.~(\ref{axionV}) and easily find that
\begin{equation}\label{massapprox}
m_i = \frac{2\pi\Lambda_i^2}{f_i} \, ,
\end{equation}
and from Eqs.~(\ref{generalinteraction}) and (\ref{axionVcorrection}),
\begin{equation}\label{couplingapprox}
g_{ij} = \frac{m_i m_j}{\sqrt{2}M_s^2} =
2\sqrt{2}\pi^2\frac{\Lambda_i^2\Lambda_j^2}{f_i f_j M_s^2} \, .
\end{equation}
Moreover, it is known~\cite{axiondecayrate} that in string theory $f_i$ is likely to
lie within
\begin{equation}
0 \lesssim f_i \lesssim m_\mathrm{Pl} \, ,
\end{equation}
so that any field value beyond $m_\mathrm{Pl}$ seems prohibited. But nevertheless as
mentioned in the previous section with a large number of fields it is possible to
have the simplest chaotic inflation with quadratic potential as
Eq.~(\ref{generalV}), which is favoured by recent cosmological observations. The
important point is that, still the predictions generally follow those of the single
field model, so that for example the overall mass scale is constrained to be
\begin{equation}\label{massconst}
\langle m \rangle \sim \mathcal{O}\left( 10^{-5} \right) m_\mathrm{Pl}
\end{equation}
from the observed amplitude of density perturbations~\cite{Pamp}. This in turn can
put a reasonable bound on the string scale $M_s$.

For simplicity, let us assume that all the masses and the couplings are the same and
that all the axion decay constants are set to be $m_\mathrm{Pl}$. The observed
density perturbations require Eq.~(\ref{massconst}), thus using
Eq.~(\ref{massapprox}) we have
\begin{equation}
m = \frac{2\pi\Lambda^2}{m_\mathrm{Pl}} \sim 10^{-5} m_\mathrm{Pl} \, ,
\end{equation}
so that the axion scale is
\begin{equation}
\Lambda \sim 10^{-3}m_\mathrm{Pl} \, .
\end{equation}
Now, using Eq.~(\ref{couplingapprox}), we can estimate
\begin{equation}
g \sim 10^{-10} \left( \frac{m_\mathrm{Pl}}{M_s} \right)^2 \, ,
\end{equation}
thus an appropriate bound on the couplings can constrain the possible range of the
string scale $M_s$: for example, from Fig.~\ref{num} let us take the bound $g
\lesssim 10^{-6}$ for the couplings not to completely dominate the dynamics. Then ,
we obtain
\begin{equation}
M_s \gtrsim 10^{-2}m_\mathrm{Pl} \sim 10^{16}\mathrm{GeV} \sim M_\mathrm{GUT} \, .
\end{equation}
Thus, in this case the string scale smaller than the scale of grand unified theory
$M_\mathrm{GUT}$ would lead to comparably short period of inflation so that it may
not solve the cosmological problems. In this context, $M_s$ as large as
$m_\mathrm{Pl}$ highly suppresses the couplings and is thus favoured.

\begin{figure}[h]
\psfrag{N}{$\mathcal{N}$}%
\psfrag{logMsmpl}{$M_s/m_\mathrm{Pl}$}%
\psfrag{logP}{$\mathcal{P}^{1/2}$}%
\begin{center}
\epsfig{file = 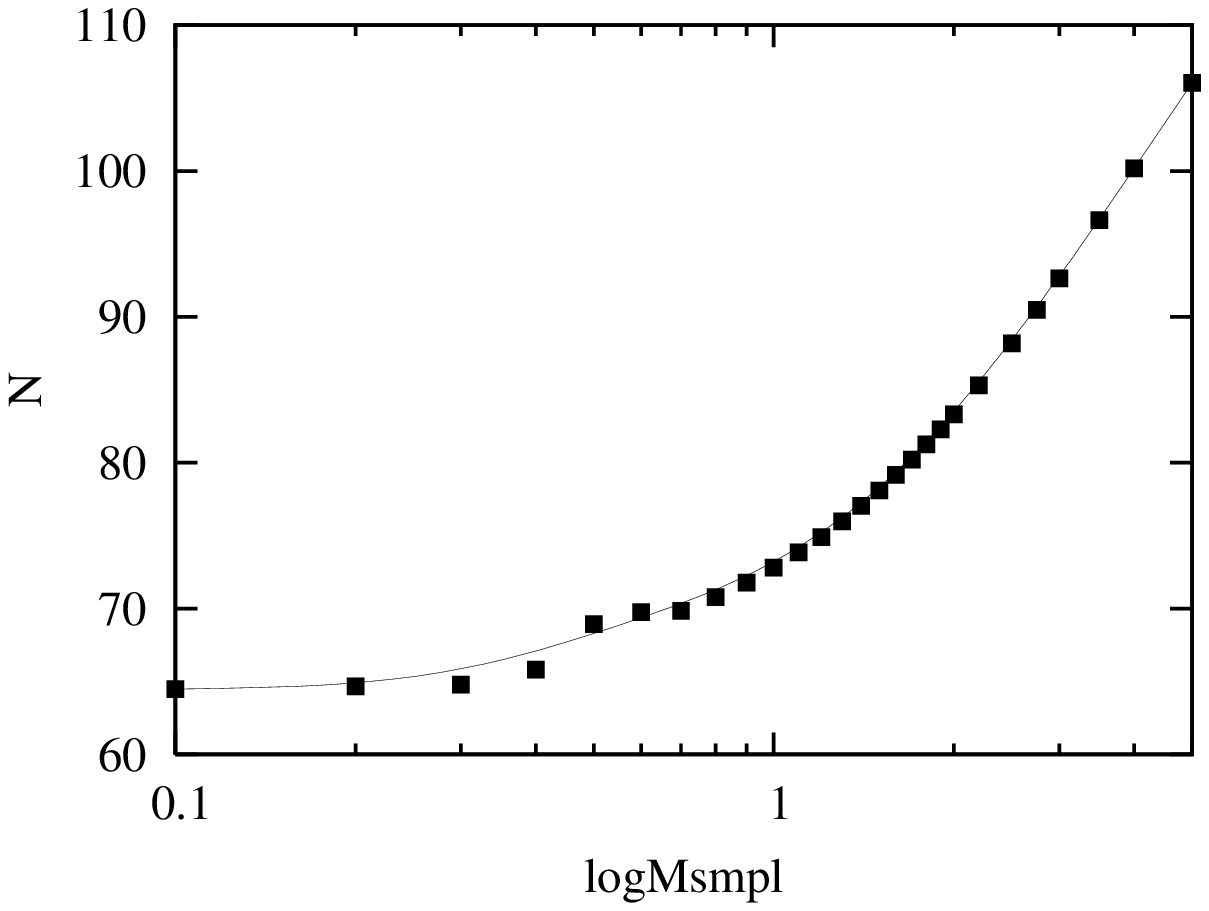, width = 6cm, angle = -90}%
\epsfig{file = 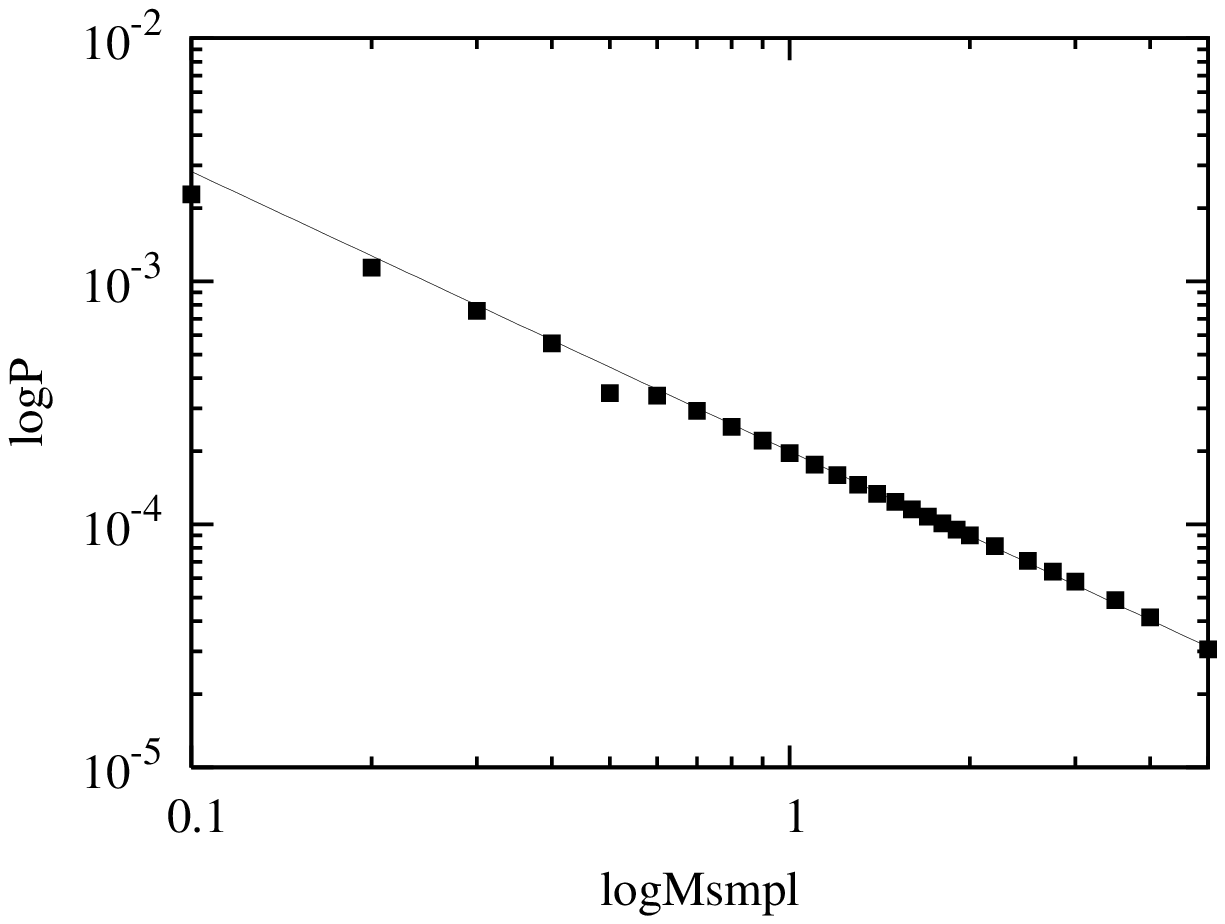, width = 6cm, angle = -90}
\end{center}
\caption{The plots of (left) $\mathcal{N}$ and (right) $\mathcal{P}^{1/2}$ versus
$M_s/m_\mathrm{Pl}$ with the same potential and parameter values. As in
Fig.~\ref{gdependence}, square dots are numerical results, and solid lines gives the
interpolated curves.}
\label{Msdependence}
\end{figure}

We can obtain the same estimate by comparing the magnitudes of Eqs.~(\ref{axionV})
and (\ref{axionVcorrection}) as follows~\cite{correction}. By assuming that all the
$\Lambda_i$ and $f_i$ are of the same order of magnitude for simplicity, we have
\begin{align}
V_\mathrm{leading} & \sim N\Lambda^4 \, ,
\\
V_\mathrm{int} & \sim \frac{(N\Lambda^4)^2}{M_s^4} \, ,
\end{align}
where we have used the fact that for a random distribution of $\theta$ between 0 and
$2\pi$, $\langle \cos^2\theta \rangle = 1/2 \sim \mathcal{O}(1)$. The interaction
energy $V_\mathrm{int}$ increases proportional to $N^2$ since it is summed over two
different indices. Since we expect that inflation is driven dominantly by
$V_\mathrm{leading}$, which should be the energy scale of inflation, we can write
\begin{equation}
\frac{V_\mathrm{leading}}{V_\mathrm{int}} \sim \frac{M_s^4}{N\Lambda^4} \sim \left(
\frac{M_s}{M_\mathrm{inf}} \right)^4 \gg 1 \, .
\end{equation}
That is, the inflationary energy scale is bounded from above by the cutoff energy
scale of the underlying theory, which is quite reasonable. To reproduce the
predictions of the chaotic inflation with quadratic potential, with the inflaton
mass given by Eq.~(\ref{massconst}), we have $M_\mathrm{inf} \sim M_\mathrm{GUT}$
about 60 $e$-folds before the end of inflation. This again gives $M_s \gtrsim
M_\mathrm{GUT}$. In Fig.~\ref{Msdependence}, we show the dependence of $\mathcal{N}$
and $\mathcal{P}$ on the string scale $M_s$.

An important point we can read from Fig.~\ref{Msdependence} is that, the coupling
$g$ {\em cannot} be indefinitely small. This is because $g$ is inversely
proportional to $M_s^2$ as Eq.~(\ref{couplingapprox}) and the current ultra violet
cutoff scale is bounded by $m_\mathrm{Pl}$ from above. We immediately face a
difficulty regarding the amplitude of $\mathcal{P}$, which becomes unacceptably
large as $M_s$ becomes even slightly less than $m_\mathrm{Pl}$ provided that the
masses are of the right magnitude which gives $\mathcal{P}^{1/2} \sim 5 \times
10^{-5}$, i.e. Eq.~(\ref{massconst}). In this case, as can be seen in the right
panel of Fig.~\ref{Msdependence}, even the largest conceivable string scale $M_s
\sim \mathcal{O}(m_\mathrm{Pl})$ is only marginally acceptable.

This estimate is, however, completely changed if we allow a post-inflationary
generation of perturbations such as the curvaton
mechanism~\cite{curvaton1,curvaton2}. Eq.~(\ref{massconst}) assumes the simplest
possibility that no curvature perturbation is generated during the post-inflationary
dynamics and that all the observed inhomogeneities in the universe are purely due to
the fluctuations of the inflaton fields. If we do not require this standard lore we
can liberate $\langle m \rangle$ from the constraint Eq.~(\ref{massconst}) and it
can be reduced by large orders of magnitude. From Eq.~(\ref{couplingapprox}) this
gives rise to a large allowed range of $M_s$. In that sense, a post-inflationary
generation of perturbation is not just an option, but is rather {\em strongly
favoured} in the light of richer string phenomenology which is due to considerably
smaller $M_s$ than $m_\mathrm{Pl}$. If additional curvature perturbation is
generated via the curvaton mechanism, it is
required~\cite{curvatoncondition1,curvatoncondition2} that during inflation $\sigma
\gtrsim 10^{-6} m_\mathrm{Pl}$, or equivalently,
\begin{equation}
H^2 \sim \frac{M_\mathrm{inf}^4}{m_\mathrm{Pl}^2} \gtrsim 10^{-22}m_\mathrm{Pl}^2 \,
,
\end{equation}
so that we are allowed have an intermediate inflationary energy scale of
$M_\mathrm{inf} \gtrsim 10^{11} \mathrm{GeV}$. Eq.~(\ref{massconst}) then can be
lowered as much as
\begin{equation}
\langle m \rangle \sim \mathcal{O}\left( 10^3 \right) \mathrm{TeV} \, .
\end{equation}
Then, from Eq.~(\ref{couplingapprox}),
\begin{equation}
g \gtrsim 10^{-24} \left( \frac{m_\mathrm{Pl}}{M_s} \right)^2 \, ,
\end{equation}
which now allows $M_s$ much lower than $M_\mathrm{GUT}$. This is perfectly
compatible with phenomenologically interesting intermediate string scales of $M_s
\sim 10^{10 - 12} \mathrm{GeV}$, which may naturally arise in specific string
compactifications such as large volume compactification~\cite{lvc}. From
Eq.~(\ref{massapprox}) this is reduced to the problem of constructing very large
$f_i$, which is still unclear in the context of string theory~\cite{axiondecayrate},
although there exist some possibilities~\cite{largedecayconst1,largedecayconst2}.

\section{Conclusions}

We have studied the dynamics of multi-field inflation in the presence of the
interactions between the inflaton fields. In single field models the inflaton field
is required to be coupled very weakly to other fields to maintain the flatness of
its potential, otherwise the successful predictions of the slow-roll conditions
which are consistent with the most recent observations are apt to get spoiled. In
multi-field models, in general all the light scalar fields contribute to the
inflationary epoch and there should exist coupled terms between the inflaton fields,
which presumably have profound effects on the inflationary dynamics. As the simplest
possibility we have taken the multi-field chaotic inflation model as the example
with quadratic coupling terms between the inflaton fields. As shown in
Fig.~\ref{num}, the first observation is that the total number of $e$-folds
$\mathcal{N}$ can be considerably reduced. This happens when the coupled terms
dominate over the bare masses, since the quadratic couplings act as the quartic
potential in the equations of motion of each inflaton field. We can follow the same
reasoning for any general power law potential and interactions, and the observable
quantities such as the power spectrum $\mathcal{P}$ and the spectral index
$n_\mathrm{s}$ vary between the predictions of the theory with the lowest power and
those of the highest power of the potential.

This consideration can place a bound on the characteristic scale of the underlying
physics such as the string scale $m_s$ for string theory. We have considered the
case where the ubiquitous string axion fields can play the role of the multiple
inflaton fields. With the assumption that they are displaced not too far from the
minimum they can reproduce the uncoupled quadratic potential, Eq.~(\ref{generalV}),
as the leading approximation, but via multi-instanton corrections different axion
fields are coupled as Eq.~(\ref{axionVcorrection}). These coupled terms are
suppressed by the string scale, thus the strength of the coupled terms is dependent
on the magnitude of the string scale. Hence from the requirement that the
interactions be not too strong to hinder a long enough period of inflation, we can
place an interesting bound on the string scale. We have considered here the simplest
quadratic potential, and in this case a large $M_s$ of at least
$\mathcal{O}(M_\mathrm{GUT})$ is preferred. But allowing post-inflationary
generation of perturbation can give much more relaxed constraint.

\subsection*{Acknowledgements}

I am grateful to Jasjeet S. Bagla, Ashoke Sen and L. Sriramkumar for comments and
suggestions. I thank the anonymous referee as well for keen criticisms on the
earlier manuscripts. It is also my great pleasure to thank the Yukawa Institute for
Theoretical Physics at Kyoto University, where this work was completed during the
YITP workshop ``Summer Institute 2007'' (YITP-W-07-12).

\end{document}